\documentclass{aip-cp}

\usepackage[numbers]{natbib}
\usepackage{rotating}
\usepackage{graphicx}

\begin{document}

\title{Dynamical Description Of The Fission Process Using The TD-BCS Theory}

\author[aff1]{Guillaume Scamps \corref{cor1}}
\author[aff2]{C\'edric Simenel}
\author[aff3]{Denis Lacroix}


\affil[aff1]{Department of Physics, Tohoku University, Sendai 980-8578, Japan}
\affil[aff2]{Department of Nuclear Physics, Research School of Physics and Engineering \\ Australian National University, Canberra, Australian Capital Territory 2601, Australia}
\affil[aff3]{Institut de Physique Nucl\'eaire, IN2P3-CNRS, Universit\'e Paris-Sud, F-91406 Orsay Cedex, France}
\corresp[cor1]{Corresponding author: scamps@nucl.phys.tohoku.ac.jp}

\maketitle

\begin{abstract}

The description of fission remains a challenge for nuclear microscopic theories. The time-dependent Hartree-Fock approach with BCS pairing is applied  to study the last stage of the fission process. A good agreement is found for the one-body observables: the total kinetic energy and the average mass asymmetry. The  non-physical  dependence of two-body observables with the initial shape is discussed.

\end{abstract}

\section{INTRODUCTION}

The fission process is an ideal phenomenon to test the predictive power of dynamical theories. Indeed, this process incorporates many aspects of nuclear dynamic, dissipation, superfluidity, tunneling, as well as a large number of degrees of freedom. The fission dynamic has been the object of several approaches: Brownian motion on the potential energy surface \cite{ran11}, time-dependent generator coordinate method (TDGCM) \cite{gou05},  time-dependent Hartree-Fock (TDHF) \cite{neg78} and adiabatic models that suppose that the two fragments choose the path that minimizes the energy from the initial well to the scission point.

The adiabatic assumption is often assumed   along the fission process, where the friction is important and so the evolution is slow enough to follow the path that minimizes the energy. Nevertheless, close to  scission, non-adiabatic effects play an important role \cite{Sim14}. A smooth transition occurs between the adiabatic motion when the two fragments are in contact and the fast  evolution due to the Coulomb repulsion when the neck is broken. This transition involves many degrees of freedom: distance between the fragments, deformations of the fragments, neck configuration ...    The TDHF theory is an  ideal tool to study this process as it  makes no restriction on the shape of the one-body density. Then the results do not depend of an arbitrary choice of parametrization of the nuclear shape during  the scission.

A limitation of the applications with TDHF is the absence of pairing. As it has been shown in References \cite{sca13,sca13b}, pairing correlation can have an important impact on dynamical processes through the initial conditions, the coupling to pair excitations, etc... 
In order to take into account pairing in the TDHF framework, the natural candidate is the time-dependent Hartree-Fock-Bogoliubov theory (TDHFB). The TDHFB theory is numerically demanding \cite{ave08}. In consequence, we choose to adopt the TD-BCS approximation. In several studies, it has been shown that the BCS approximation is a good compromise to take into account pairing in the mean-field framework. In particular, comparisons to  QRPA calculations show a good agreement with the BCS approximation for the small amplitude motion \cite{sca13b,eba10}.

\section{THE FISSION PROCESS}

In order to study the fission process, the system is first initialized after the barrier on the adiabatic path. This initialization is done using the Constraint Hartree-Fock with BCS approximation (CHF+BCS). A modified version of the code EV8 \cite{bon05} has been used, where two of the reflection symmetry have been removed allowing the octupole deformation necessary to study asymmetric fission. As a test case, we  study the  $^{258}$Fm fission where experimental data are available  \cite{hul89}. In the literature, three modes are considered: a symmetric fission with compact fragments, a symmetric fission with elongated fragments and an asymmetric fission. The three modes exhibit different dynamical behaviors   due to different structure effects and Coulomb energy at the scission point.

\begin{figure}[h]
  \centerline{\includegraphics[width=450pt]{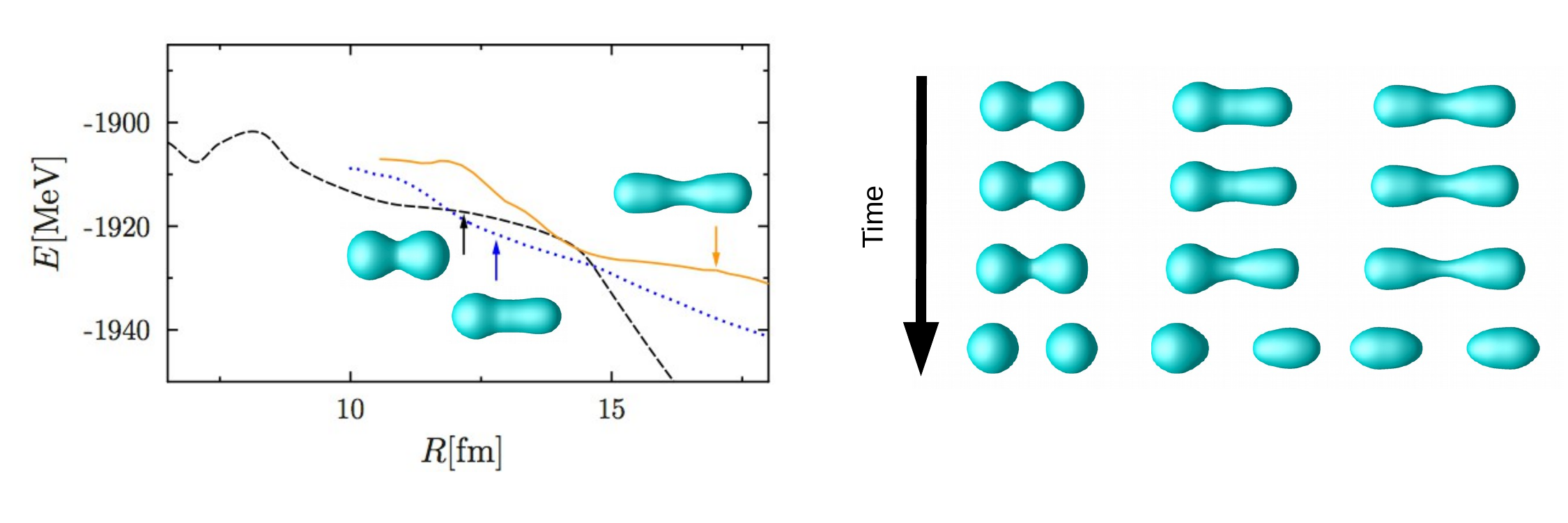}}
  \caption{ Left panel: Energy as a function of the distance between the two fragments for the three modes. The arrows represent the starting configuration of the dynamical calculation. Right panel: Isosurface density as a function of time for the three modes. The time between two pictures is different for the three modes, from left to right, $\Delta t=$ 0.675 zs, 1.8 zs and 1.08 zs.  } \label{fig:pot}
\end{figure}

Using the CHF+BCS, the evolution of the potential energy is shown on Figure \ref{fig:pot} (left). Each mode is associated with a valley in the 3 dimensional space ($Q_{20}$, $Q_{30}$, $Q_{40}$) and is shown here as a function of the distance between the two fragments. Similar results for the potential energy as a function of the deformation for the three modes have been found using different interactions or methods in References \cite{bon06,sta09}. Starting from the three valleys, three corresponding TD-BCS evolutions have been done. The evolutions until complete scission are shown on the Figure \ref{fig:pot} (right). The details of the calculation can be found in Reference \cite{sca15}.

\subsection{One-body observables}

\begin{figure}[h]
  \centerline{\includegraphics[width=450pt]{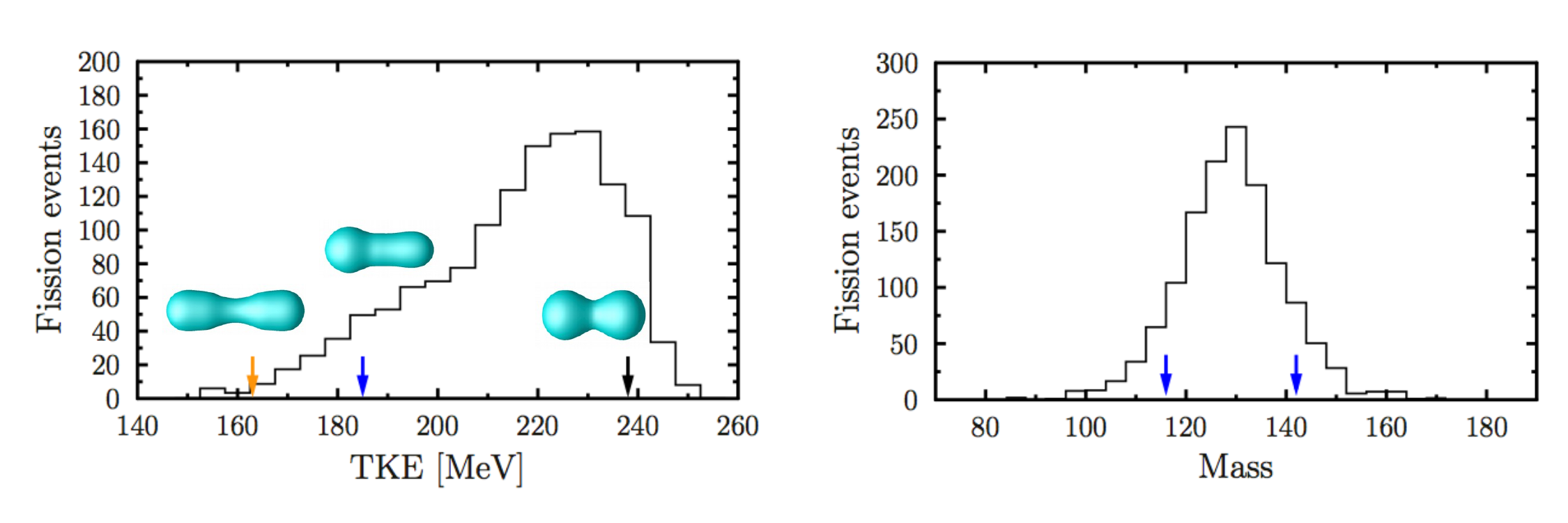}}
  \caption{ Left panel:  Comparison between the TKE obtained for the three mode to the experimental TKE distribution. Right panel:  Comparison between the average mass in the asymmetric mode to the experimental mass distribution (only events with TKE $<$ 220 MeV are shown).  } \label{fig:TKE}
\end{figure}

Several observables can be extracted from this simulation of the fission process. In particular, the total kinetic energy (TKE) is calculated after the complete separation of the two fragments by adding the Coulomb energy to the kinetic energy. A comparison between the TD-BCS and the experimental data is shown in Figure \ref{fig:TKE}. The TD-BCS can predict neither the population of each of the mode nor the fluctuation of the TKE. Nevertheless, it predicts an average TKE and mass distribution for each mode in good agreement with the experimental distributions.  For the elongated mode, it is found that this mode is compatible with the tail of the TKE distribution. 

We can worry about the dependence of those observables with the starting point of the dynamic. In a naive picture, starting with a more compact shape  increases the total energy and should then increase the TKE. In reality, due to the strong friction forces, all this additional energy is dissipated during the dynamic. For example, for initial distance of the fragment between 9.5 and 12.5 fm (both well before scission), the TKE is the changed by less than 1 MeV for the symmetric  compact mode. The fact that this observable is independent of the starting point confirms that the evolution between $R$= 9.5 fm and 12.5 fm is adiabatic.

\subsection{Two-body observables}

\begin{figure}[h]
  \centerline{\includegraphics[width=450pt]{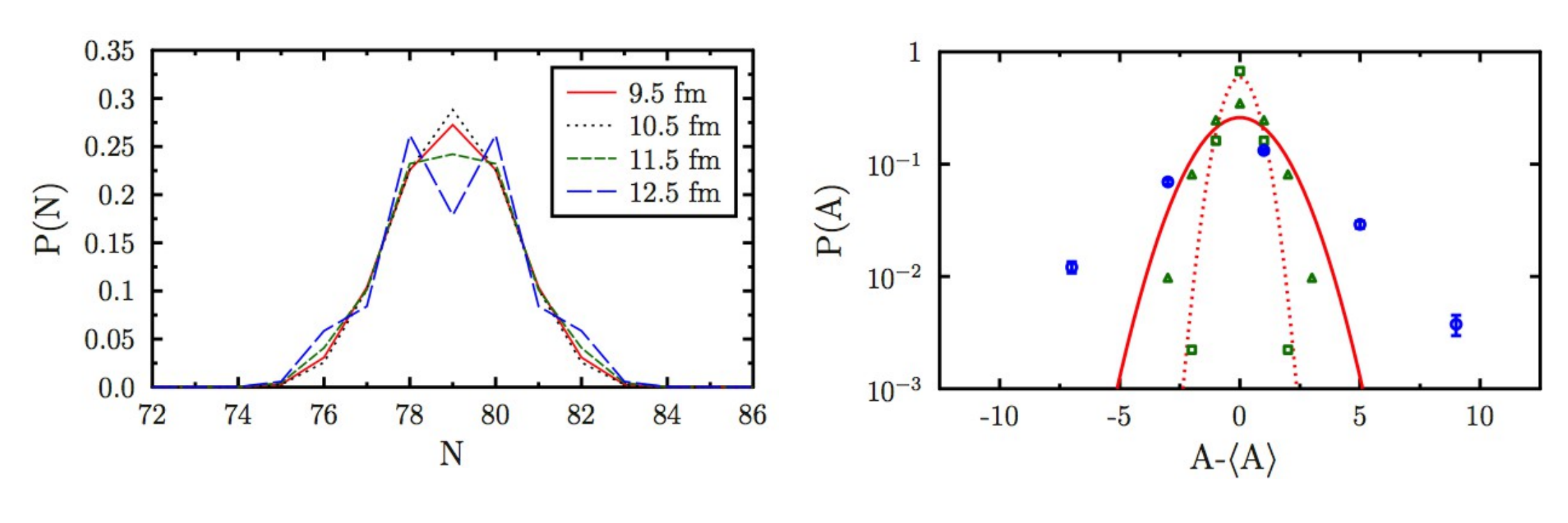}}
  \caption{ Left panel:  Neutron distribution of the symmetric compact mode for different initial distances between the fragments. Right panel:  Mass distribution obtained with TDHF (for  $^{264}$Fm) with initial distance $R=$10 fm (triangles) and $R=$ 12 fm (squares). The TDRPA results are shown by solid line for $R=$10 fm and dotted line for $R=$12 fm assuming a Gaussian distribution. The experimental mass distribution (for $^{258}$Fm) is also shown with blue circles.  } \label{fig:Pair_width}
\end{figure}

The two-body observables, like the  odd-even effects or the fluctuations of the mass distribution show a different behavior. To study this effect, we display on Figure \ref{fig:Pair_width} (left), the neutron distribution with different initial distances between the fragments. The distribution is obtained using the projection technique developed for TDHF in Reference \cite{sim10b} and extended to the case with pairing in Reference \cite{sca13}.

Starting from a configuration close to the scission point ($R=$12.5 fm), the neutron distribution shows an odd-even effects  due to the pairing correlations. When more compact initial shapes are chosen, the odd-even effects are smoothed out. This effect can be understood simply by the fact that the initial energy is larger for compact shape. Then more energy is dissipated during the descent of the potential. This dissipated energy break the pair correlation and so the odd-even effects are reduced.

We next investigate the impact of the initial configuration on the width of the fragment mass distribution in Figure \ref{fig:Pair_width} (right). The fluctuations of the mass distribution is obtained without pairing with the TDHF theory and is compared to fluctuation obtained with the time-Dependent Random Phase Approximation (TDRPA) theory \cite{bal84,sim11}. The mass distribution is shown in Figure \ref{fig:Pair_width} (right) assuming a gaussian shape. The TDRPA goes beyond the TDHF approach with a variational approach not only for the one-body observables but also for the fluctuations. The TDRPA result is then closer to the experimental data. Nevertheless, for the two theories a dependence of the results with the initial distance is found. 

This is expected as the fluctuations accumulate along the entire fission path from the compound nucleus to scission. We can expect that a dynamical theory starting from the initial well would reproduce the experimental fluctuation. Nevertheless as shown in References \cite{god15,Tan15}, within mean-field dynamics theories, the fission process does not occur for too compact shape.

The role of initiale excitation should also be investigate. One could to take into account the excitation energy in the initial quasi-particle vacuum state by doing a finite temperature calculation. Indeed, the first part of the dynamic being  slow enough for the excitation energy to be thermalized during the fission process.

\section{CONCLUSION}

The fission process has been study with the TD-BCS theory. It is shown that the one-body observables does not depends of the initial adiabatic configuration and reproduce the experimental data. For the two-body observables, a dependence is found. The latter could be due to the non-consideration of the initial excitation energy in the dynamical calculations. This dependence shows the necessity to go beyond the present approach with finite temperature calculations.

\section{ACKNOWLEDGMENTS}
 G.S. acknowledges the Japan Society for the Promotion of Science
 for the JSPS postdoctoral fellowship for foreign researchers.
 This work was supported by Grant-in-Aid for JSPS Fellows No. 14F04769. 
This work has been supported by the
Australian Research Council Grants No. FT120100760.


\nocite{*}
\bibliographystyle{aipnum-cp}%
\bibliography{biblio2}%

\end{document}